# Influence of the linear magneto-electric effect on the lateral shift of light reflected from a magneto-electric film

**Yu S Dadoenkova**[1,2,3], **F F L Bentivegna**[4], **N N Dadoenkova**[2,3], **R V Petrov**[1], **I L Lyubchanskii**[3] **and M I Bichurin**[1]

[1]Institute of Electronic and Information Systems, Novgorod State University, Veliky Novgorod 173003, Russia
[2]Ulyanovsk State University, Ulyanovsk 432017, Russia
[3]Donetsk Physical and Technical Institute of the NAS of Ukraine, Donetsk 83114, Ukraine
[4]Lab-STICC (UMR 6285), CNRS, UBL, École Nationale d'Ingénieurs de Brest, 29238 Brest Cedex 3, France

**Abstract.** We present a theoretical investigation of the lateral shift of an infrared light beam reflected from a magnetic film deposited on a non-magnetic dielectric substrate, taking into account the linear magneto-electric interaction in the magnetic film. We use the stationary phase method to evaluate the lateral shift. It is shown that the magneto-electric coupling leads to a six-fold enhancement of the lateral shift amplitude of a $p$-($s$-) polarized incident beam reflected into a $s$-($p$-) polarized beam. A reversal of the magnetization in the film leads to a nonreciprocal sign change of the lateral shift.

## 1. Introduction

When a spatially limited optical beam impinges on an optical system, the reflected beam may be laterally shifted in the plane of incidence with respect to the behaviour predicted by geometric optics for idealized light rays. This effect, called the Goos-Hänchen (GH) effect [1], can be observed in many configurations and optical structures, and has recently found practical applications, e.g. for the design of an optical heterodyne sensor [2] or an optical waveguide switch [3], for biosensing technologies [4] or for the detection of chemical vapours [5]. The GH shift is very sensitive to the nature and geometrical characteristics of the constituents of the optical system and thus can be used to measure such parameters as refractive indices, beam apertures and incidence angles, temperature, displacement, or film thicknesses [2].

The development of magneto-photonic devices, however, has also raised the issue of the potential influence of magnetization on the GH shift. In magneto-optical materials, the lateral shift has been shown to reach values up to several tens of light wavelengths [6-8] and to be tunable *via* the application of an external magnetic field [9, 10]. Thus the GH shift ought to be taken into account for the design of integrated optical or magneto-optical devices.

It is also well known that some magnetic materials possess magneto-electric (ME) properties [11-13], which result in the induction of a magnetization by an electric field or induction of a dielectric polarization by a magnetic field. Since it has been shown that the linear ME interaction can greatly affect the complex reflectivity of a magnetic/dielectric bilayer [14-16], it can thus be expected that ME properties also influence the GH shift of a light beam upon reflection from a system including magnetic materials.

This paper purports to study the dependency of the GH shift on the ME interaction in a magnetic film deposited on a non-magnetic dielectric substrate, and to determine how this shift could be controlled through the application of an external magnetic field to that system.

**2. General analysis**

Let us consider the reflection of a finite-size light beam from a magnetic film of thickness $d$ deposited on a semi-infinite dielectric isotropic substrate, as depicted in figure 1. The interfaces between the materials are parallel to the ($xy$) plane of a Cartesian system of coordinates. An electromagnetic harmonic wave of angular frequency $\omega$ impinges the upper surface of the film under oblique incidence angle $\theta$ from the vacuum. Without loss of generality, the plane of incidence can be chosen as ($xz$) and the incident and reflected electromagnetic waves can be decomposed into $s$- and $p$- components of their electric field strengths $E_{s,p}^{(i,r)}$ with respect to that plane, where superscripts ($i$) and ($r$) correspond to the incident and reflected fields, respectively. The magnetic film is magnetized at saturation in the polar magneto-optical configuration, i. e., perpendicularly to the interface, and its magnetization can be reversed *via* an externally-applied magnetic field. We assume that this magnetic film exhibits linear ME coupling.

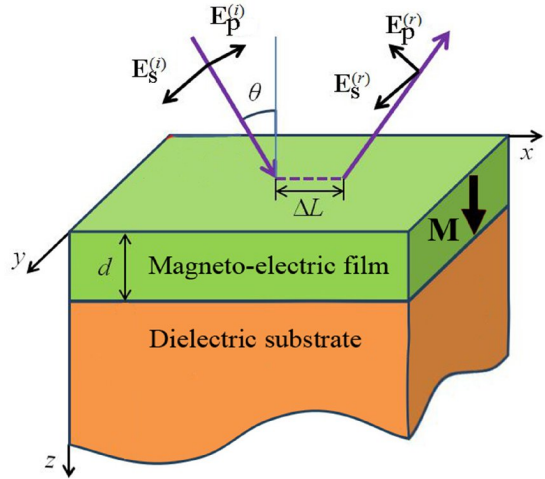

**Figure 1.** Schematic of the system under study: a magnetic film (thickness $d$) grown on a semi-infinite isotropic dielectric substrate. The saturation magnetization $\mathbf{M} = \{0, 0, M_s\}$ in the magnetic film is perpendicular to the plane of the film. The $s$- and $p$-components of the incident ($i$) and reflected ($r$) optical electric fields are denoted $E_{s,p}^{(i,r)}$. The GH shift in the plane of incidence upon reflection is $\Delta L$.

When a wavepacket impinges on the uppermost surface of the magnetic film, its reflection is known to depart from the behaviour predicted by ray optics, in that it undergoes a non-negligible GH lateral shift $\Delta L$ in the plane of incidence (in our case, along the $x$-axis in figure 1). Note that this shift can assume positive as well as negative values. A shift towards increasing values of $x$, such as represented in figure 1, will thus be deemed positive. The amplitude $\Delta L$ of the lateral shift can be obtained with the stationary-phase approach initially proposed by Artmann [17]. Assuming a monochromatic incident Gaussian beam of waist $w_0$ whose transverse field distribution is given by

$$E^{(i)}(x) = E_0 \exp\left[-\frac{x^2}{2w_0^2}\right], \quad (1)$$

the lateral shift of the reflected wavepacket then simply writes [7]

$$\Delta L = -\frac{\partial \psi}{\partial k_x} + \frac{\partial \ln |R|}{\partial k_x} \frac{\partial^2 \psi}{\partial k_x^2} \left(w_0^2 + \frac{\partial^2 \ln |R|}{\partial k_x^2}\right)^{-1}, \quad (2)$$

where $R(k_x)$ is the complex reflection coefficient for any given $k_x$ component of the incident wavevector $\mathbf{k}$ along the $x$-axis.

Taking into account the linear ME interaction, the constitutive material equations expressing the relationship in the frequency domain of the electric displacement vector $\mathbf{D}$ and the magnetic induction $\mathbf{B}$ in the magnetic film with the electric field $\mathbf{E}$ and the magnetic field $\mathbf{H}$ of the electromagnetic wave read [11]:

$$D_i = \varepsilon_0 \varepsilon_{ij} E_j + \alpha_{ij} H_j, \quad B_i = \mu_0 \mu_{ij} H_j + \alpha_{ij} E_j, \quad (3)$$

where $\varepsilon_0$ and $\mu_0$ are the vacuum permittivity and permeability, $\varepsilon_{ij}$ and $\mu_{ij}$ are the relative permittivity and permeability tensor elements of the magnetic film, and $\alpha_{ij}$ are the elements of the ME tensor of the magnetic medium. The latter is diagonal in crystals with cubic symmetry ($\alpha_{ij} = \alpha \, \delta_{ij}$) [11].

The numerical simulations described in Section 3 will be carried out with a magnetic garnet as a magnetic medium. Indeed, magnetic garnets, which are widely used in magneto-photonic devices, exhibit good transparency properties in the near-infrared regime. They are also magnetically bigyrotropic in that regime, i.e., their relative permittivity and permeability tensors both depend on the local magnetization. Hence, tensors $\varepsilon_{ij}$ and $\mu_{ij}$ can be expanded in power series of the magnetization vector $\mathbf{M}$ [18]. Neglecting terms above the first-order terms in such expansions, they write, in the polar magneto-optical configuration and in optically isotropic crystals:

$$\varepsilon_{ij} = \begin{pmatrix} \tilde{\varepsilon} & if^{(e)}m_z & 0 \\ -if^{(e)}m_z & \tilde{\varepsilon} & 0 \\ 0 & 0 & \tilde{\varepsilon} \end{pmatrix}, \quad \mu_{ij} = \begin{pmatrix} \tilde{\mu} & if^{(m)}m_z & 0 \\ -if^{(m)}m_z & \tilde{\mu} & 0 \\ 0 & 0 & \tilde{\mu} \end{pmatrix}, \quad (4)$$

where $m_z = M_z/|\mathbf{M}|$, $\tilde{\varepsilon}$ and $\tilde{\mu}$ are the crystallographic components of the diagonal relative permittivity and permeability tensors of the medium, and $f^{(e)}$ and $f^{(m)}$ are the linear gyroelectric and gyromagnetic coefficients of the crystal.

Taking into account the boundary conditions at each interface of the system (in our geometry, continuity of the tangential $x$- and $y$-components of the electric and magnetic fields), one can then relate the amplitudes of the reflected $E_{s,p}^{(r)}$ and incident $E_{s,p}^{(i)}$ optical electric field components *via* the reflection matrix $\hat{\mathfrak{R}}$ as:

$$\begin{pmatrix} E_s^{(r)} \\ E_p^{(r)} \end{pmatrix} = \hat{\mathfrak{R}} \begin{pmatrix} E_s^{(i)} \\ E_p^{(i)} \end{pmatrix}, \quad \text{with} \quad \hat{\mathfrak{R}} = \begin{pmatrix} \mathfrak{R}_{ss} & \mathfrak{R}_{sp} \\ \mathfrak{R}_{ps} & \mathfrak{R}_{pp} \end{pmatrix}. \quad (5)$$

In general, the reflection matrix $\hat{\mathfrak{R}}$ has four non-zero components. Its off-diagonal components correspond to cross-polarization upon reflection of light and arise from the optical anisotropy of magneto-optical and ME couplings and thus, in turn, from the crystalline symmetry of the magnetic film.

Using Eq. (2) and taking into account Eqs. (3-5), one can finally obtain the value of the GH shift for each of the four polarimetric combinations of incident and reflected field components. In the following Section, these lateral shifts will be presented in the reduced form $\Delta X_{ij} = \Delta L_{ij}/\lambda_0$ ($i, j = s, p$), where $\lambda_0$ is the incident light wavelength. The reduced shift $\Delta X_{ps}$ (resp., $\Delta X_{sp}$) thus describes the cross-polarized contribution to the GH effect, i.e., the lateral shift of a $s$- (resp., $p$-) polarized incident wave reflected into a $p$- (resp., $s$-) polarized wave.

## 3. Results and discussion

The calculation of the lateral shift in the system described in Section 2 was performed in the case of a magnetic layer of yttrium iron garnet (YIG) $Y_3Fe_5O_{12}$ grown on a substrate of gadolinium gallium $Gd_3Ga_5O_{12}$ garnet. Symmetry rules forbid the linear ME effect in crystals with an inversion centre such as YIG. However, in thin epitaxial films (of thicknesses less than 4 μm) symmetry lowering takes place, and the inversion centre vanishes [19]. The linear ME constant in thin YIG films can reach large values up to 30 ps.m$^{-1}$ [19]. For our numerical calculations, we considered a magnetic film thickness $d = 2$ μm and used the following material parameters (at $\lambda_0 = 1.15$ μm): $\tilde{\varepsilon} = 4.56$, $\tilde{\mu} = 1$, $f^{(e)} = -2.47 \times 10^{-4}$, $f^{(m)} = 8.65 \times 10^{-5}$ [20], and $\alpha = 30$ ps.m$^{-1}$ [19].

In order to establish the influence of the ME coupling on the optical properties of the system, hence on the GH shift it imposes to an incident Gaussian beam, we have compared the dependency of the latter as a function of the beam incidence angle $\theta$ when the ME interaction is taken into account in our calculations ($\alpha = 30$ ps.m$^{-1}$) and when it is not ($\alpha = 0$). Our results first show that the GH shift observed upon the reflection of a $s$- ($p$-)polarized incident wave into a similarly $s$- ($p$-)polarized wave almost does not depend on the ME interaction. On the other hand, the ME interaction noticeably affects the cross-polarized lateral shift upon the reflection of a $p$-($s$-) polarized incident wave into a $s$-(-$p$) polarized wave.

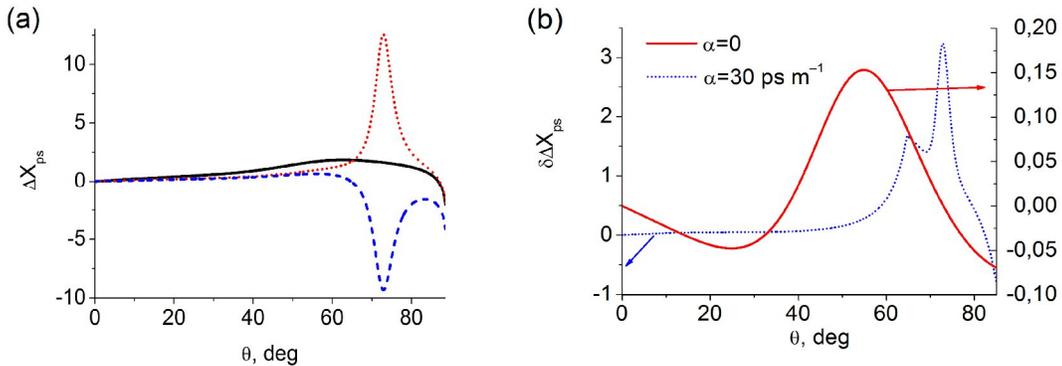

**Figure 2.** (a) Reduced GH shift $\Delta X_{ps}$ as a function of the incidence angle $\theta$ of light for $\alpha = 0$ (ME coupling neglected, black solid line) and $\alpha = 30$ ps.m$^{-1}$ (ME coupling taken into account) for $M_s > 0$ ($\Delta X_{ps}(M \uparrow\uparrow z)$, red dotted line) and for $M_s < 0$ ($\Delta X_{ps}(M \uparrow\downarrow z)$, blue dashed line); (b) Difference in the GH shift $\delta\Delta X_{ps} = \Delta X_{ps}(M \uparrow\uparrow z) - |\Delta X_{ps}(M \uparrow\downarrow z)|$ upon magnetization reversal for $\alpha = 0$ (red solid line) and $\alpha = 30$ ps.m$^{-1}$ (blue dotted line).

The reduced GH shift $\Delta X_{ps}$ is shown in figure 2(a) as function of the incidence angle $\theta$. Comparing the black solid and red dotted lines, one can see that $\Delta X_{ps}$ exhibits up to a six-fold increase when the ME coupling is taken into account. This enhancement is maximal for incidence angles close to about 73°, where $\Delta X_{ps}$ reaches a peak value of about $12\lambda_0$. In this case, the magnetization **M** in the film points towards the overall direction of the incident beam ($M_s > 0$ and $k_z > 0$), as shown in figure 1, and the shift is positive. It should be noted that both cross-polarized contributions to the GH shift are equal ($\Delta X_{ps} = \Delta X_{sp}$) due to the crystalline symmetry of the magnetic crystal YIG.

Moreover, using an external magnetic field to reverse the initial magnetization **M** of the magnetic film inverts the GH shift direction [see red dotted and blue dashed lines in figure 2(a)] in the presence of ME interaction in the magnetic film. In this case the amplitude of the peak of the lateral shift is about $-9\lambda_0$, and its angular position experiences a slight displacement of about 1° towards larger values of $\theta$. The difference in $\Delta X_{ps}$ upon magnetization reversal is shown in figure 2(b). One observes that the inversion of the GH shift is non-symmetrical when the ME coupling is taken into account ($\alpha = 30$ ps.m$^{-1}$), as the difference $\delta\Delta X_{ps} = \Delta X_{ps}(M \uparrow\uparrow z) - |\Delta X_{ps}(M \uparrow\downarrow z)|$ is non-zero [blue dotted line in figure 2(b)] and can reach up to a value of almost 3. Neglecting the ME coupling in the calculations ($\alpha = 0$), on the other hand, results only in a very slight change of the GH effect upon magnetization reversal [red solid line in figure 2(b)], without inversion of the direction of the lateral shift. This confirms that the ME effect does play an essential part in the GH shift in the system.

This conclusion is further substantiated with a study of the effect of the value of the ME constant $\alpha$ on the GH shift, as shown in figure 3 for both directions of saturation magnetization in the magnetic layer. As already seen in figure 2(b), there is almost no change in the GH shift upon magnetization reversal for $\alpha = 0$. When $\alpha$ gradually increases, however, the contrast in GH shift between the two opposite directions of saturation magnetization keeps increasing as well, and both the sign reversal of the shift and the asymmetry of that reversal (i.e., with $\delta\Delta X_{ps} \neq 0$) become more pronounced. Overall, the maximum value of the shift increases (in absolute value) with $\alpha$. For small values of $\alpha$, an increase of that constant leads to a slight shift of the maximum of $\Delta X_{ps}$ towards larger incidence angles for both directions of the magnetization [compare the black solid, red dotted, and blue dashed lines in figure 3]. For increasingly larger values of $\alpha$, the angular position of that maximum remains almost unchanged [see blue dashed and magenta dash-dotted lines in figure 3], both upon magnetization reversal and upon an increase of the ME constant.

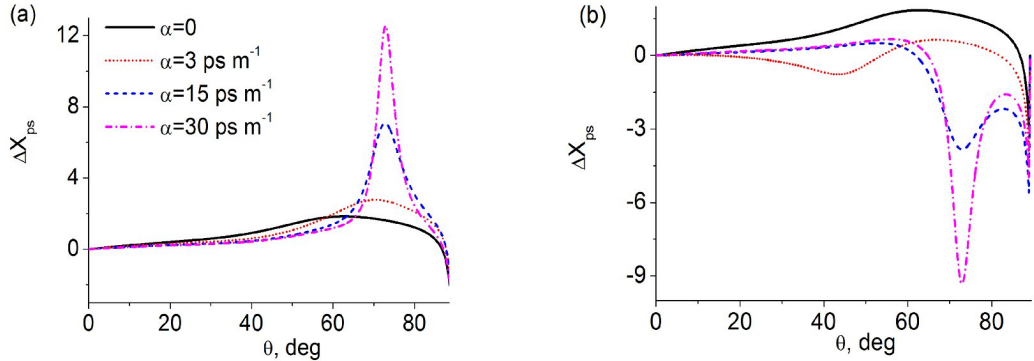

**Figure 3.** Reduced GH shift $\Delta X_{ps}$ as a function of the incidence angle $\theta$ of light for different values of the ME constant $\alpha = 0$, 3, 15, and 30 ps.m$^{-1}$ (black solid, red dotted, blue dashed, and magenta dash-dotted lines, respectively) for **(a)** $M \uparrow\uparrow z$ and **(b)** $M \uparrow\downarrow z$.

Consequently, this sensitivity of the GH shift to $\alpha$ makes it possible to use GH shift measurements in order to estimate the ME constant of a magneto-electric medium (provided it is transparent at the wavelength used for measurements), in contrast to Faraday rotation measurements [15]. More generally, the value and sign of the GH shift are sensitive to all the materials properties mathematically summarized by Eqs. (3) and (4), upon which the reflectivity factors of Eq. (5) depend in a complex but entirely analytical way. Measuring the lateral shift of light for a set of well-chosen angles of incidence and states of polarization of the incident and reflected beams can be a way to detect small variations of the material properties under the influence of external conditions (strain, wave frequency, temperature, etc.). Moreover, the ability to control, enhance, reduce or reverse the GH shift in such a magneto-electric heterostructure provides a route towards magnetically-controlled photonic devices.

## 4. Conclusions

We have studied theoretically the lateral, or Goos-Hänchen, shift of an infrared light beam reflected from the upper surface of a magnetic film deposited on a non-magnetic dielectric substrate, taking into account the linear magneto-electric interaction in the magnetic film. It has been shown that the magneto-electric coupling leads to an about six-fold increase of the lateral shift upon reflection of a *p*-(*s*-) polarized incident wave to a *s*-(*p*-) polarized wave, up to about 10 times the incident wavelength. Magnetization reversal leads to a nonreciprocal sign change of the shift when magneto-electric coupling is present in the system.

Thus, the linear magneto-electric interaction, in combination with an external magnetic field, allows to controllably enhance the lateral shift of light to values that can be easily measured. This shows that the use of magneto-electric materials can find applications, through the measurement of the Goos-Hänchen shift they induce, for instance to the determination of very tiny variations of material properties, or to the design of magnetically-tunable optical switches and sensors.

**Acknowledgments**
This research is supported partly by a grant from the Russian Science Foundation (Project No. 15-19-10036); a grant from the Ministry of Education and Science of the Russian Federation (Project No. 14.Z50.31.0015); European Union's Horizon 2020 research and innovation programme under the Marie Skłodowska-Curie grant agreement (Project No. 644348); the MPNS COST Action (Project No. MP1403); grant No. 14MOESMOMO14 from École Nationale d'Ingénieurs de Brest (France).